\documentclass[aps,prl,twocolumn]{revtex4}

\usepackage{amssymb,amsmath,graphicx}

\begin{document}

\title{Impact of non-Poisson activity patterns on spreading processes}

\author{Alexei Vazquez$^1$, Bal\'azs R\'acz$^2$, Andr\'as Luk\'acs$^2$
and Albert-L\'aszl\'o Barab\'asi$^3$\\
$^1$ The Simons Center for Systems Biology, Institute of Advanced Study,
Einstein Drive, Princeton, NJ 08540, USA\\
$^2$Computer and Automation Research Institute, Hungarian Academy of
Sciences (MTA SZTAKI), Budapest, Hungary\\
$^3$Department of Physics and Center for Complex Networks Research,
University of Notre Dame, IN 46556, USA}

\date{\today}

\begin{abstract}

Halting a computer or biological virus outbreak requires a detailed 
understanding of the timing of the interactions between susceptible and 
infected individuals. While current spreading models assume that users 
interact uniformly in time, following a Poisson process, a series of 
recent measurements indicate that the inter-contact time distribution is 
heavy tailed, corresponding to a temporally inhomogeneous bursty contact 
process. Here we show that the non-Poisson nature of the contact dynamics 
results in prevalence decay times significantly larger than predicted by 
the standard Poisson process based models. Our predictions are in 
agreement with the detailed time resolved prevalence data of computer 
viruses, which, according to virus bulletins, show a decay time close to a 
year, in contrast with the one day decay predicted by the standard Poisson 
process based models.

\end{abstract}

\maketitle

\bibliographystyle{apsrev}


According to The WildList Organization International (www.wildlist.org)
there were 130 known computer viruses in 1993, a number that has exploded
to 4,767 in April 2006. With the proliferation of broadband ``always on''
connections, file downloads, instant messaging, Bluetooth-enabled mobile
devices and other communications technologies the mechanisms used by worms
and viruses to spread have evolved as well. Still, most viruses continue
to spread through email attachments. Indeed, according to the Virus
Bulletin (www.virusbtn.com), the Email worms W32/Netsky.h and W32/Mytob
with the ability to spread itself through email, account for 70\% of the
virus prevalences in April 2006. When the worm infects a machine, it sends
and infected email to all addresses in the computer's email address book.
This self-broadcast mechanism allows for the worm's rapid reproduction and
spread, explaining why email worms continue to be the main security
threat.

In order to eradicate viruses, as well as to control and limit the impact
of an outbreak, we need to have a detailed and quantitative understanding
of the spreading dynamics. This is currently provided by a wide range of
epidemic models, each adopted to the particular realities of the computer
based spreading process. A common feature of all current epidemic models
\cite{anderson91,pv00,mpv02,meyers04,bbv04,moreno04,nevokee05,boccaletti06}
is the assumption that the contact process between individuals follows
Poisson statistics, meaning that the probability that an agent interacts
with another agent in a $dt$ time interval is $dt/\langle\tau\rangle$,
where $\langle\tau\rangle$ is the mean interevent time. Furthermore, the
time $\tau$ between two consecutive contacts is predicted to follow an
exponential distribution with mean $\langle\tau\rangle$. Therefore,
reports of new infections should decay exponentially with a decay time of
about a day, or at most a few days
\cite{anderson91,pv00,mpv02,meyers04,bbv04}, given that most users check
their emails on a daily basics, providing $\langle\tau\rangle$ of
approximately a few days (see below). In contrast, prevalence records
indicate that new infections are still reported years after the release of
antiviruses (http://www.virusbtn.com,\cite{pv00,pv04}), and their decay
time is in the vicinity of years, two-three orders of magnitude larger
than the Poisson process predicted decay times.

A possible resolution of this discrepancy may be rooted in the failure of
the {\it Poisson approximation} for the inter-event time distribution,
currently used in all modeling frameworks. Indeed, recent studies of email
exchange records between individuals in a university environment have
shown that the probability density function $P(\tau)$ of the time interval
$\tau$ between two consecutive emails sent by the same user is well
approximated by a fat tailed distribution $P(\tau)\sim \tau^{-1}$
\cite{eckmann04,johansen04,barabasi05,vazquez05b,vazquez06d,vazquez06g}.
In the following we provide evidence that this deviation from the Poisson
process has a strong impact on the email worm's spread, offering a
coherent explanation of the anomalously long prevalence times observed for
email viruses.

\begin{figure}[t]
\centerline{\includegraphics[width=3in]{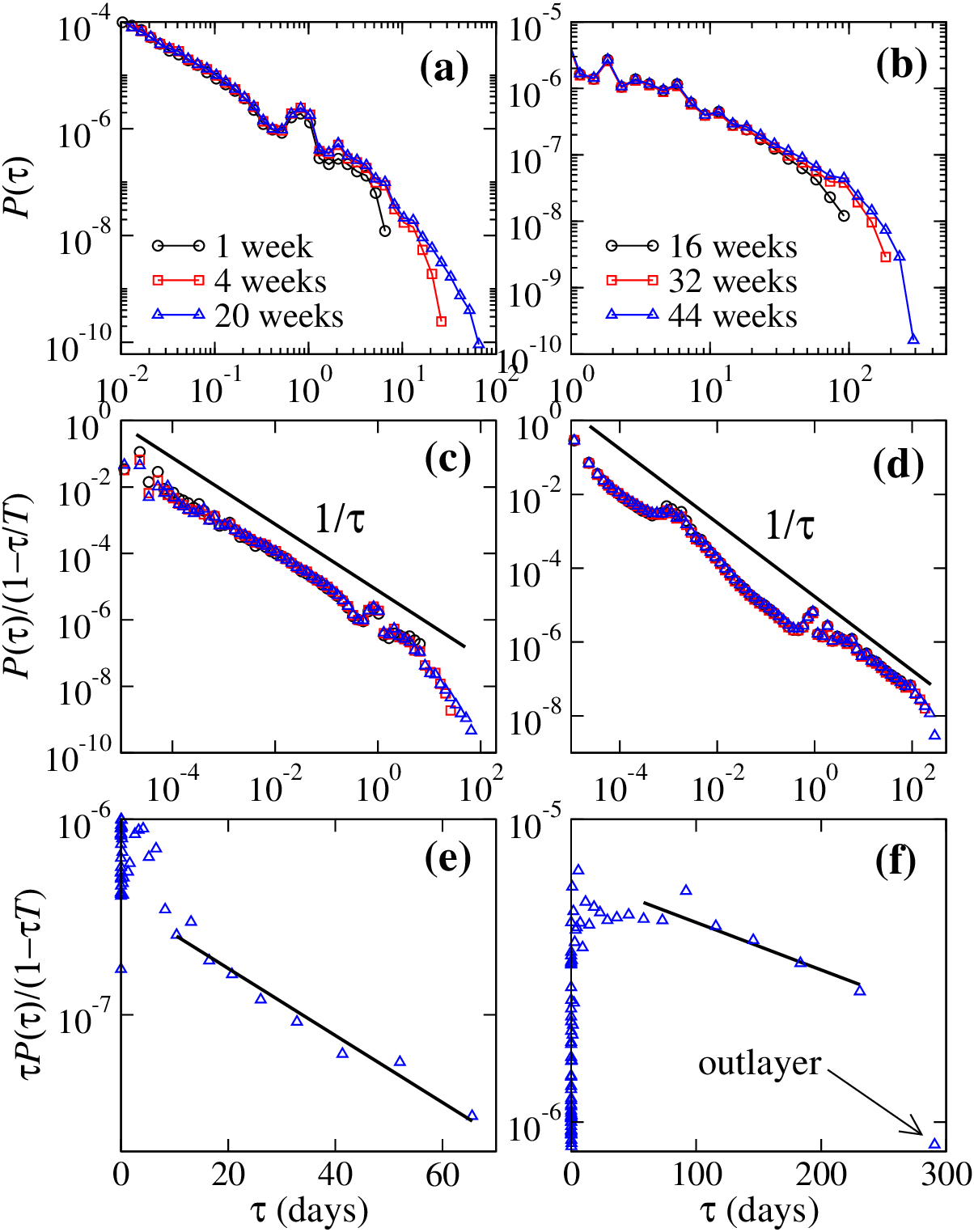}}

\caption{Distribution $P(\tau)$ of the inter-event time between two
consecutive emails sent by an email user. The left and right panels
represent the university and commercial datasets, respectively. For each
dataset we aggregate the interevent times of all users (the distribution
for single users exhibits a similar behavior \cite{vazquez06d}) and apply
a logarithmic binning to account for the fact that the number of observed
events decrease with increasing $\tau$. (a)  and (b)  Log-log plot of
$P(\tau)$ for $\tau>10^{-2}$ days, emphasizing the large $\tau$ behavior
for different time window sizes $T$. (c)  and (d) The same plots after
removing the finite time window effects, the data collapsing into a single
curve. The solid line represents the power law decay
$P(\tau)\sim\tau^{-1}$. (e) and (f) Semilog plot emphasizing the
exponential decay at large $\tau$, for the largest time window $T$. The
solid lines are fit to an exponential decay $\tau P(\tau)\sim
e^{-\tau/\tau_{\rm E}}$ resulting in $\tau_{\rm E}=25\pm2$ days and
$9\pm1$ months for the university (e) and commercial (f) datasets,
respectively. The outlayer in (f) was excluded when fitting to an
exponential decay.}

\label{fig1}
\end{figure}

{\it Email activity patterns:} The contact dynamics responsible for the
spread of email worms is driven by the email communication and usage
patterns of individuals. To characterize these patterns we studied two
email datasets. The first dataset contains emails from a university
environment, capturing the communication pattern between 3,188 users,
consisting on 129,135 emails \cite{eckmann04}. The second dataset contains
emails from a commercial provider (freemail.hu) spanning ten months,
1,729,165 users and 39,046,030 emails. For the two email datasets
$P(\tau)$ is rather broad, following approximately a power law with
exponent $\alpha\approx1$ and a cutoff at large $\tau$ values (Fig.
\ref{fig1}). Most important, the value of the cutoff depends on the time
window $T$ over which the data has been recorded (Fig. \ref{fig1}a,b). By
restricting the data to varying time windows we find that $P(\tau)$ goes
to zero as $1-\tau/T$ when $\tau$ approaches $T$. After correcting for the
finiteness of the observation time window we obtain that the distributions
for different $T$ values collapse into a single curve (Fig.
\ref{fig1}c,d), representing the true inter-event time distribution. The
obtained $P(\tau)$ is well approximated by a power law decay followed by
an exponential cutoff (Fig. \ref{fig1}c-f), i.e.

\begin{equation}
P_{\rm E}(\tau)=A\tau^{-\alpha}
\exp\left(-\frac{\tau}{\tau_{\rm E}}\right)\ ,
\label{PtauEmail}
\end{equation}

\noindent where $A$ is a normalization factor. The power law decay at
small and intermediate $\tau$ is clearly manifested on the log-log plot of
$P(\tau)$ (Fig. \ref{fig1}c,d), consistent with $\alpha\approx1$, spanning
over four (Fig. \ref{fig1}c) to six (Fig. \ref{fig1}d) decades. The
exponential cutoff is best seen in a semi-log plot after removing the
power law decay (Fig. \ref{fig1}e,f), resulting in a decay time $\tau_{\rm
E}=25\pm2$ days and $\tau_{\rm E}=9\pm1$ months (approximately 270
days) for the university and commercial datasets, respectively (see
Fig. \ref{fig1}c-f). In contrast, the Poisson approximation predicts
$P_{\rm P}(\tau)=\exp(-t/\langle\tau\rangle)/\langle\tau\rangle$
\cite{fellerII}, where $\langle\tau\rangle$ is the mean interevent time,
taking the values 0.86 and 4.9 days for the university and commercial
data, respectively.

{\it The dynamics of worm spreading:} To investigate the impact of the
observed non-Poisson activity patterns on spreading processes we study the
spread of email worms among email users.  For the moment we ignore the
possibility that some users may delete the infected email or may have
installed the worm antivirus and therefore do not participate in the
spreading process, to return later at the possible impact of these events
on our predictions. Therefore, the spreading process is well described by
the susceptible-infected (SI) model on the email network.

The spreading dynamics is jointly determined by the email activity
patterns and the topology of the corresponding email communication network
\cite{emb02,eckmann04}. The email activity patterns are reflected in the
infection generation times, where the generation time is defined as the
time interval between the infection of the primary case (the user sending
the email) and the infection of a secondary case (a different user opening
the received infected email). From the perspective of the secondary case,
the time when a user receives the infected email is random and the
generation time is the time interval between arrival and the opening of
the infected email. In most cases received emails are responded in the
next email activity burst \cite{eckmann04,vazquez06d}, and viruses are
acting when emails are read, approximately the same time when the next
bunch of emails are written. Therefore the generation time can be
approximated by the time interval between the arrival of a virus infected
email, and the next email sent to any recipient by the secondary case. If
we model the email activity pattern as a renewal process \cite{fellerII}
with inter-event time distribution $P(\tau)$ then the generation time is
the residual waiting time and is characterized by the probability density
function \cite{fellerII}

\begin{equation}
g(\tau) = \frac{1}{\langle\tau\rangle} \int_\tau^\infty dx P(x)\ .
\label{gtau}
\end{equation}

Next we calculate the average number of new infections $n(t)$ at time $t$
resulting from an outbreak starting from a single infected user at $t=0$.
Although the email network contain cycles, it is a very sparse, thus we
approximate it by a tree-like structure. Previous analytical studies have
shown that this approximation captures the main features of the spreading
dynamics on real networks \cite{pv00,vazquez06b}. In this case $n(t)$ is
given by \cite{vazquez06b}

\begin{equation}
n(t) = \sum_{d=1}^D z_d g^{\star d}(t)\ ,
\label{nt}
\end{equation}

\noindent where $z_d$ is the average number of users $d$ email contacts
away from the first infected user, $D$ is the maximum of $d$ and $g^{\star
d}(t)$ is the $d$-order convolution of $g(\tau)$, $g^{\star 1}(t)=g(t)$
and $g^{\star d}(t) = \int_0^t d\tau g(\tau)g^{\star d-1}(t-\tau)$ for
$d>1$, representing the probability density function of the sum of $d$
generation times. Substituting the Poisson approximation and Email data
interevent time distributions into (\ref{gtau}) and the result into
(\ref{nt}) we obtain

\begin{equation}
n(t) = F(t) \exp\left( -\frac{t}{\tau_0} \right)\ ,
\label{ntexp}
\end{equation}

\noindent where $\tau_0=\langle\tau\rangle$ for the Poisson approximation
and $\tau_0=\tau_{\rm E}$ for the Email data, and

\begin{equation}
F(t) = \left\{
\begin{array}{ll}
\frac{1}{\langle\tau\rangle} \sum_{d=1}^D \frac{z_d}{(d-1)!}
\left( \frac{t}{\langle\tau\rangle} \right)^{d-1}\ , &
\mbox{Poisson approx.}\\
\sum_{d=1}^D z_d f^{\star d}(t)\ , &
\mbox{Email data}\ ,
\end{array}
\right.
\label{Ft}
\end{equation}

\noindent where $f(t) = \int_\tau^\infty dx x^{-\alpha} 
e^{(tau-x)/\tau_E}/\langle\tau\rangle$. In the long time limit 
(\ref{ntexp}) is dominated by the exponential decay while $F(t)$ gives 
just a correction. The decay time is, however, significantly different for 
the Poisson approximation and the real inter-event time distribution.

\begin{figure}[t]
\centerline{\includegraphics[width=2in]{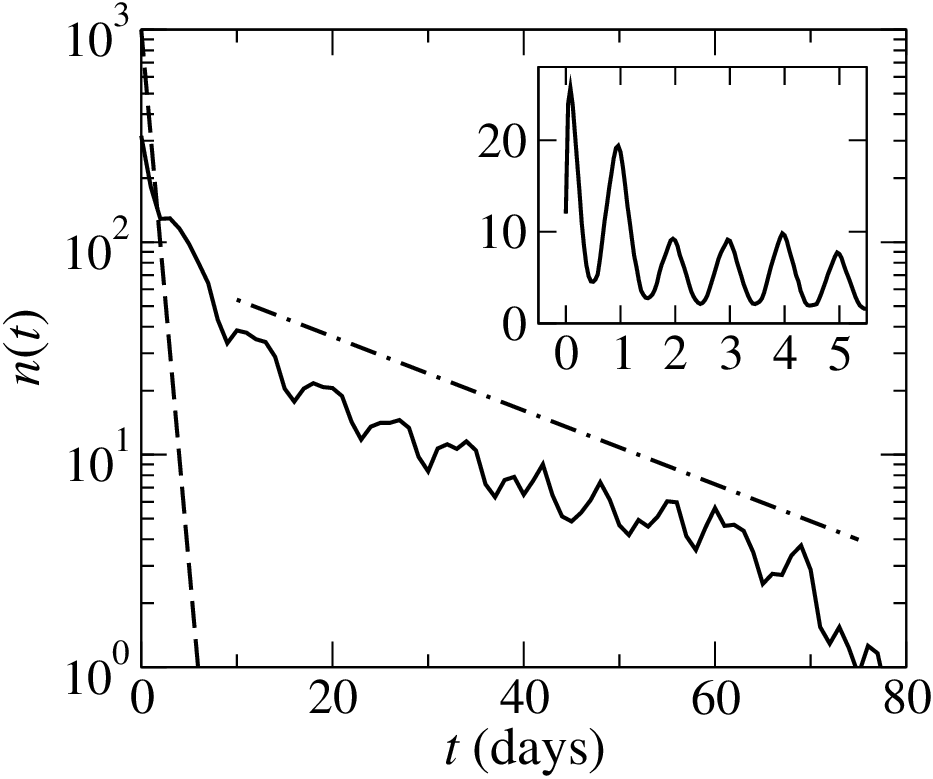}}

\caption{Average number of new infections resulting from simulations using
the email history of the university dataset (solid line), using a one day
interval binning. The inset shows a zoom of the initial stages of the
spreading process using a one hour interval binning. The lines correspond
to the exponential decay predicted by the Poisson process approximation
(dashed) and the true inter-event distribution (dot-dashed).}

\label{fig2}
\end{figure}

To test these predictions we perform numerical simulations using the
detailed email communication history. In this case a susceptible user
receiving an infected email at time $t$ becomes infected and sends an
infected email to all its email contacts at $t^\prime>t$, where $t^\prime$
is the time he/she sends an email for the first time after infection, as
documented in the email data. To reduce the computational cost we focus
our analysis on the smaller university dataset. The average number of new
infected users resulting from the simulation exhibits daily (Fig.
\ref{fig2}, inset) and weekly oscillations (Fig.  \ref{fig2}, main panel),
reflecting the daily and weekly periodicity of human activity patterns.
More important, after ten days the oscillations are superimposed on an
exponential decay, with a decay time about 21 days (see Fig. \ref{fig1}b).
The Poisson process approximation would predict a decay time of one day,
in evident disagreement with the simulations (Fig. \ref{fig2}). In
contrast, using the correct inter-event time distribution for the
university dataset we predict a decay time of $25\pm2$ days, in good
agreement with the numerical simulations (Fig. \ref{fig2}).


\begin{figure}[t]
\centerline{\includegraphics[width=3in]{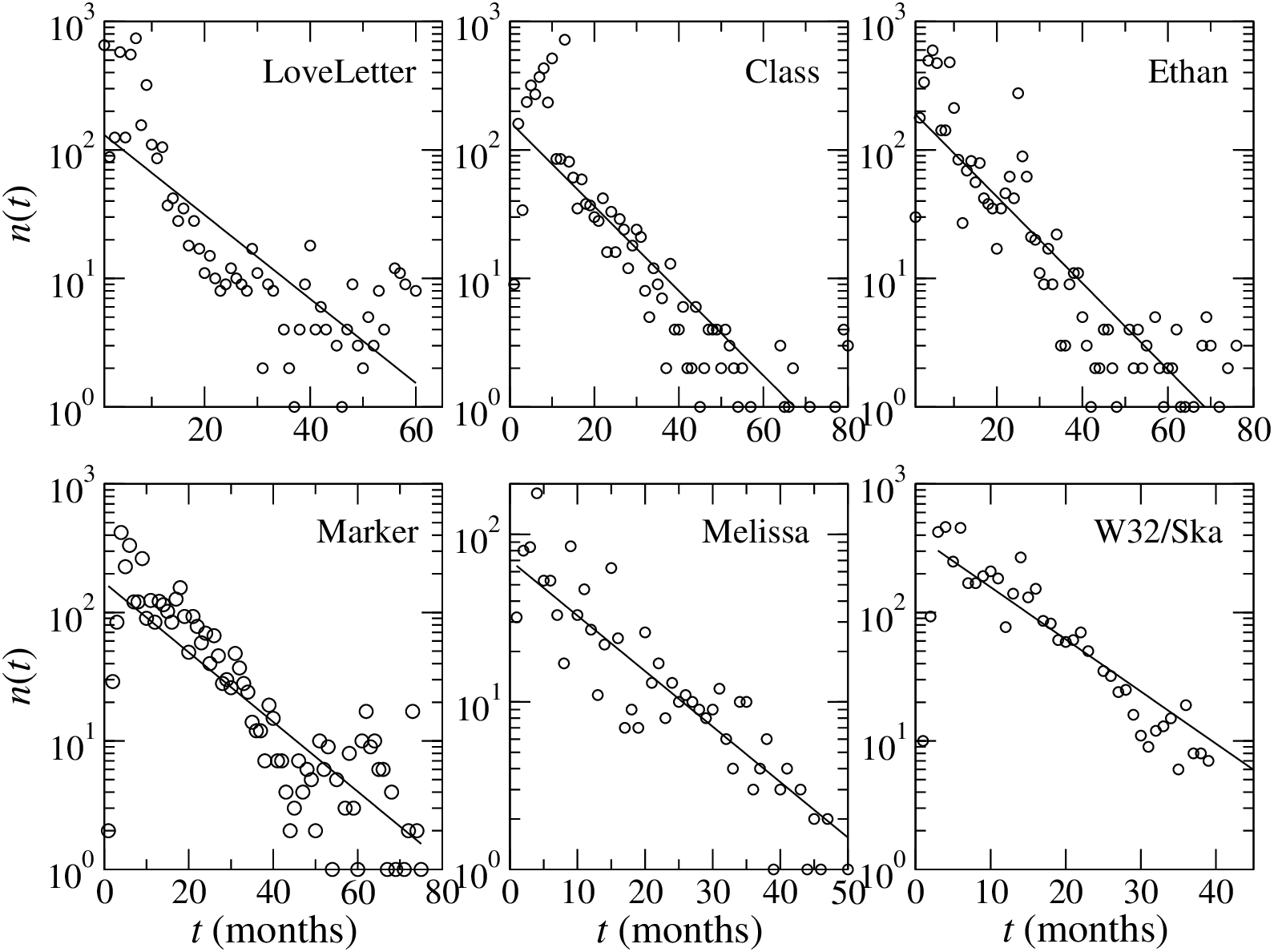}}

\caption{Number new infections reported for six worm outbreaks, according
to Virus Bulletin (www.virusbtn.com). The lines are fit to an exponential
decay resulting in the decay times (measured in months): LoveLetter
($13\pm2$), Ethan ($12\pm1$), Marker ($14\pm2$), Class ($12\pm1$), Melissa
($13\pm1$), W32/Ska ($11\pm1$).}

\label{fig3}
\end{figure}

The analysis of the university dataset allows us to demonstrate the
connection between the long $\tau$ behavior of the inter-event time
distribution $P(\tau)$ and the long time decay of the prevalence $n(t)$.
Our main finding is that the prevalence decay time is given by the
characteristic decay time of the inter-event time distribution. More
important, we show that the Poisson process approximation clearly
underestimates the decay time. For Poisson processes the two time scales,
the average interevent time and the characteristic time of the exponential
decay coincide, being both of the order of one to at most a few days.
Using measurements on the commercial dataset, containing a larger number
of individuals and covering a wider spectrum of email users, we can
extrapolate these conclusions to predict the behavior of real viruses.
Given the value of $\tau_{\rm E}$ for the commercial dataset we predict
that the email worm prevalence should decay exponentially with time, with
a decay time about nine months. The prevalence tables reported by the
Virus Bulletin web site (http://www.virusbtn.com) indicate that worm
outbreaks persist for several months, following an exponential decay with
a decay time around twelve months (Fig. \ref{fig3}). Our nine month
prediction is thus much closer to the observed value than the
$\langle\tau\rangle\approx1\div4$ day prediction based on the Poisson
approximation. The fact that our prediction underestimates the actual
decay time by about three months is probably rooted in the fact that the
commercial dataset, despite its coverage of an impressive 1.7 million
users, still captures only a small segment (approximately 0.1\%) of all
Internet users.

As we discussed above, some other factors potentially affecting the
spreading of email worms were not considered in our analysis. First, some
users may delete the infected emails or may have installed the worm
antivirus. Since these users do not participate in the spreading process
they are eliminated from the average number users $z_d$ that are found $d$
email contacts away from the first infected user. While this would affect
the initial spread characterized by $F(t)$ (\ref{Ft}), the exponential
decay in (\ref{ntexp}) and the decay time $\tau_0=\tau_{\rm E}$
will not be altered. Second, some email viruses do not use the
self-broadcasting mechanism of email worms. For example, file viruses
require the email user to attach the infected file into a sent email in
order to be transmitted. In turn, only some email contacts will receive
the infected file. Once again, this affects $z_d$ but not the email
activity patterns. Therefore, the prevalence of email viruses in general
should decay exponentially in time with a decay time $\tau_0$ determined
by the decay time of the inter-event time distribution $\tau_{\rm E}$.
Third, new virus strains regularly emerge following small modifications of
earlier viruses. Within this work new virus strains are modeled as new
outbreaks. An alternative approach is to analyze all strains together,
modeling the emergence of new strains as a process of reinfection. In this
second approach the dynamics is better described by the
susceptible-infected-susceptible (SIS) model \cite{pv00}. Earlier work has
shown that if reinfections are allowed in networks with a power law degree
distribution, long prevalence decay times may emerge, which increase with
increasing the network size \cite{pv00}. The data shown in Fig. \ref{fig3}
represent, however, the spread of a single virus strain, which is better
captured by the SI model. For the SI model, however, for a Poisson
activity pattern we should get a rapid decay in prevalence, indicating
that the empirically observed long decay times cannot be attributed to
this reinfection-based mechanism.

A series of recent measurements indicate that power law inter-event time
distributions are not a unique feature of email communications, but emerge
in a wide range of human activity patterns, describing the timing of
financial transactions \cite{plerou00,masoliver03}, response time of
internauts \cite{johansen01}, online games \cite{henderson01}, login times
into email servers \cite{chatterjee03} and printing processes
\cite{harder06}. Together they raise the possibility that non-Poisson
contact timing are a common feature of human dynamics and thus could
impact other spreading processes as well. Indeed, measurements indicate
that the patterns of visitation of public places, like libraries
\cite{vazquez06d}, or the long range travel patterns of humans, involving
car and air travel, is also driven by fat tailed inter-event times
\cite{brockmann06}. Such travel patterns play a key role in the spread of
biological viruses, such as influenza or SARS \cite{colizza06}. Taken
together, these results indicate that the anomalous decay time predicted
and observed for email viruses may in fact apply more widely, potentially
impacting the spread of biological viruses as well.

{\bf Acknowledgements:} We wish to thank Deok-Sun Lee for useful comments
on the manuscript. This work was supported by a grant from the James
McDonell Foundation and NSF, by a Yahoo Faculty Research Grant and grant
\emph{ASTOR} NKFP 2/004/05, and the Mobile Innovation Center, Hungary.


\end{document}